\documentclass[%
 reprint,
 amsmath,amssymb,
 aps,
]{revtex4-1}

\usepackage{graphicx}
\usepackage{dcolumn}
\usepackage{bm}

\usepackage{color}
\usepackage{colortbl}

\usepackage{xcolor}

\def \bea{\begin{eqnarray}}
\def \eea{\end{eqnarray}}

\def \c2V{ \cos { 2 \theta_{hel}(D^*)}}

\def \mm2 {M^{2}_{M}}

\def \cosDstar{\cos \theta_{\rm hel}^{D^*}}

\def \M2miss {$M^{2}_{\rm miss}$}

\newcommand{\anti}[1]
{
	\overline{#1}\mbox{}
}

\begin{document}

\title{Semitauonic $B$ decays at Belle/Belle II}

\author{Karol~Adamczyk}
\affiliation{%
Institute of Nuclear Physics Polish Academy of Sciences, Krakow
}
\collaboration{On behalf of the Belle/Belle II Collaboration}

\date{\today}

\begin{abstract}

It is experimentally observed that ratios of branching fractions for semitauonic and semileptonic $B$ decays, known as the $R(D^{(*)})$, are higher than Standard Model (SM) predictions. The $B \to \anti{D}^{(*)} \tau^{+} \nu_{\tau}$ decays, except for the branching fractions, offer also other observables, in particular, angular observables may determine spin structure of a potential New Physics (NP) providing some clues to explain the $R(D^{(*)})$ puzzle. In this review, 
polarization measurements in $B \to \anti{D}^{*} \tau^{+} \nu_{\tau}$
decays\footnote{Charge-conjugate modes are implied throughout this raport, unless explicitly stated.} at Belle are  presented and prospects for measurements of characteristics of semitauonic $B$ decays at Belle/Belle II are described.

\end{abstract}

\maketitle

\section{Experimental situation}

Semitauonic $B$ decays are described as $b \to c \tau \nu$ transitions that occure at the tree level and in SM are mediated by the exchange of virtual $W^*$ boson. Presence of the $3^{rd}$ generation massive fermions in the initial and final states makes them sensitive to potential effects of NP. In particular, the relative rates, 
\begin{equation}
R(D^{(*)})=\frac{\mathcal{B}(B \to \anti{D}^{(*)}\tau^+\nu_{\tau})}
{\mathcal{B}(B \to \anti{D}^{(*)}\ell^+\nu_{\ell})} \quad (\ell = e, \mu)
\end{equation} 
are useful observables to test lepton flavour universality (LFU), since the most of theoretical (the $|V_{cb}|$ element of the CKM matrix and some hadronic form factors) and experimental uncertainties (reconstruction efficiencies, particles identification) can be reduced.

Analyses of $B$ meson decays with one or more undetected particles in the final states are challenging, due to poor signal signature. However, at the $e^+ e^-$ environment, where $B$ mesons are produced exclusively in pairs, reconstruction of one $B$ meson, called tagging $B$ ($B_{\rm tag}$), provides information on the momentum vector and other quantum numbers of the signal side ($B_{\rm sig}$).
For semitauonic $B$ decays the most efficient tagging technique is inclusive reconstruction. In the above approach signal decays ($B_{\rm sig}$) are reconstructed first, and then accompanying $B$ meson ($B_{\rm tag}$) is reconstructed inclusively from all the particles that remain after selecting the $B_{\rm sig}$. Subsequent, efficient way of reconstruction of the tag side is using frequent and clean semileptonic $B$ decays ($ \Gamma(B \to \anti{D}^{(*)}\ell\nu_{\ell}) \sim 20\%$), though without giving precise constraint on the momentum vector of $B_{\rm tag}$. 
Complete kinematic information on the tag side with high purity of sample can be obtained by combining many exclusive $B$ decay modes in, so called, full reconstruction of events (FR). Purity of the sample is maximized at the price of lower efficiency, however, final efficiency and purity depend on the particular analysis. 
Various tagging methods mean (partially) independent subsets of data sample, but what's more, they provide cross-checks of results, due to different sources and proportions of systematic uncertainties for measured observables.


Exclusive semitauonic $B$ decays were first observed in 2007~\cite{pub-dtaunu}, and since then the Babar and Belle collaborations performed a few analyses to measure $R(D)$ and $R(D^{*})$ using different
methods for the reconstruction of tagging side~\cite{Lees:2012xj, chargedHiggs, pub-dtaunu2, Huschle, Sato, Hirose,Hirose2}. 
LHCb collaboration has measured $R(D^{*})$ and similar ratios with $J/\psi$ in place of $D^{*}$~\cite{LHCb_R_dtaunu_mu,LHCb_R_dtaunu_pi,LHCb_R_dtaunu_Jpsi}.
The world averages of the most current measurements of $\mathcal{R}(D^{(*)})$~\cite{HFLAV} are 
\begin{equation}
\begin{split}
R(D) = 0.407 \pm 0.039 \pm 0.024 \\
R(D^{*}) = 0.306 \pm 0.013 \pm 0.007
\end{split}
\end{equation}
and exceed the SM predictions~\cite{HFLAV} 
\begin{equation}
\begin{split}
R(D) = 0.299 \pm 0.003 \\
R(D^{*}) = 0.257 \pm 0.003 
\end{split}
\end{equation}
by 2.3 and 3.0 standard deviations ($\sigma$) respectively. Considering the $R(D)-R(D^{*})$ correlation of -0.203, the tension with the SM predictions corresponds to about 3.78$\sigma$. Interestingly, it is also difficult to accommodate the observed branching fractions within two Higgs doublet models (2HDM)\cite{chargedHiggs}, mainly due to the relatively large excess in the $B \to \anti{D}^{*} \tau \nu$ mode, which is expected to be less sensitive to the charged Higgs contributions than the $B \to \anti{D} \tau \nu$ channel. 
So far, measured kinematic distributions like $q^2$(squared mometum transfer corresponding to the mass squared of the virtual $W^*$ decaying to the $\tau^{+}\nu_{\tau}$ system)~\cite{Huschle}, lepton~\cite{Lees:2012xj,Sato} and $D^{(*)}$ momenta~\cite{pub-dtaunu2, Sato} are consistent with the SM predictions, however still not significant to discriminate NP scenarios or to explain the the $R(D^{(*)})$ puzzle. 
Thus, an effective way to use the experimentally available statistics would be the study of integrated distributions or relevant observables, like lepton or vector meson polarizations.

\section{\label{sec:level}Polarization measurements in $B \to \anti{D}^{*} \tau^{+} \nu_{\tau}$}

The kinematics of the $B \to \anti{D}^{*} \tau^{+} \nu_{\tau}$ decay can be expressed in terms of the following variables: $q^2$, two angles $\theta_{l}$ and $\chi$, describing the primary $B$ decay, and two or more angles describing $D^*$ and $\tau$ decays~\cite{Datta}.   
$\theta_{l}$ is the angle between the $\tau$ and direction opposite to the $D^{*}$ meson momenta in $W^*$ rest frame;
$\chi$ is the azimuthal angle between two decay planes spanned by the momentum vectors of $W^*$ ($\tau^{+}\nu_{\tau}$) and $D^*$ ($D \pi$) systems; 
$\theta_{\rm hel}^{D^*}$ is the angle between the $D$ meson momentum and direction opposite to $B$ in $D^*$ rest frame.
In case of 2-body $\tau$ decays e.g. $\tau \to \pi \nu_{\tau}$,
$\theta_{\rm hel}^{\tau}$ is the angle between $\tau$ daughter from two body decay and direction opposite to $W^*$ in $\tau$ rest frame.
At B-factories the reconstruction of 
$B_{\rm tag}$ in hadronic modes allows for determination the squared transfer momentum: $q^2 = (p_{\rm B_{\rm sig}} - p_{D^{*}})^2$, where $\vec{p}_{\rm B_{\rm sig}} = - \vec{p}_{\rm B_{\rm tag}}$ and $E_{\rm sig}$ equals beam energy ($E_{\rm beam}$) in the rest frame of $\Upsilon(4S)$.
Among listed angles, two helicity angles $\theta_{\rm hel}^{\tau}$ and $\theta_{\rm hel}^{D^*}$
can be reconstructed with optimal resolution using hadronic decays of $B_{\rm tag}$ and used for
measurements of $\tau$ and $D^*$ polarizations.

Next subsections present analyses dedicated to polarization measurements in semitauonic $B$ decays that are based on the full data sample of 772 $\times 10^6$ $B\bar{B}$ pairs accumulated with the Belle detector at the $\Upsilon$(4S) resonance in the $e^+e^-$ asymmetric collider KEKB.

\subsection{$\tau$ polarization measurement in $B \to D^{*} \tau \nu$}

Using FR tagging method and two-body $\tau$ decays $\tau^+ \to \pi^+ \anti{\nu}_{\tau}$ 
and $\tau^+ \to \rho^+ \anti{\nu}_{\tau}$, the first measurement of the $\tau$ lepton polarization ($P_{\tau}$) simultaneously with $R(D^*)$ has been made\cite{Hirose, Hirose2} for combined
$B^0 \to \anti{D}^{(*)} \tau^{+} \nu_{\tau}$ and $B^{+} \to \anti{D}^{(*)0} \tau^{+} \nu_{\tau}$ decays.
The $\tau$ lepton polarization is defined as: 
\begin{equation*}
P_{\tau}=\frac{\Gamma^{+} - \Gamma^{-}}
{\Gamma^{+} + \Gamma^{-}},
\end{equation*}
where $\Gamma^{\pm}$ denotes the decay rate of $B \to \anti{D}^{*} \tau^{+} \nu_{\tau}$ with $\tau$ helicity of $\pm1/2$.
The SM value of $\tau$ polarization in $B \to \anti{D}^{*} \tau^{+} \nu_{\tau}$ decay is $P_{\tau} = -0.497\pm0.013$\cite{Tanaka}, however it can be significantly modified by NP.
%
In two-body $\tau$ decays ($\tau \to h \nu_{\tau}$) $P_{\tau}$ can be extracted from 
the following formula:   
\begin{equation*}
\frac{d\Gamma}{d\cos { \theta_{\rm hel}^{\tau}}} = \frac{1}{2} (1 + \alpha P_{\tau}\cos { \theta_{\rm hel}^{\tau}}), 
\end{equation*},
%
%
%
%
%

where $h = \pi$ the coefficient $\alpha =1$, and for $h = \rho$,  $\alpha = 0.45$.
%

Measurement of $\cos { \theta_{\rm hel}^{\tau}}$ distribution is demanding because it is modified by cross-feeds from signal events with other $\tau$ decays, and background contamination.
To measure $P_{\tau}$, the region of $\cos { \theta_{\rm hel}^{\tau}}$ is
divided into two bins: $\cos { \theta_{\rm hel}^{\tau}} > 0$ (forward) and $\cos { \theta_{\rm hel}^{\tau}} < 0$ (backward). 
The value of $P_{\tau}$ is then extracted from the forward-backward asymmetry of the signal yields, and it is given by the formula:
\begin{equation*}
P_{\tau} = \frac{2}{\alpha} \frac{ N_{\rm{sig}}^{F} - N_{\rm{sig}}^{B} }{ 
 N_{\rm{sig}}^{F} + N_{\rm{sig}}^{B} }, 
\end{equation*}
where the superscript $F(B)$ denotes the signal yield in the forward (backward) region.

For the $\tau \to \pi\nu$ mode the region of $\cos { \theta_{\rm hel}^{\tau}} > 0.8$ is excluded 
from the analysis due to a large peaking background coming from $B \to \anti{D}^* \ell^{+} \nu_{\ell}$ decays. 
Corrections to the raw $P_{\tau}$ value are applied to take into account detector effects (acceptance, asymmetric $\cos { \theta_{\rm hel}^{\tau}}$ bins, crosstalks between different $\tau$ decays).  
In the presented analysis, the value of $P_{\tau}$ is measured simultaneously with $R(D^*)$. 
The number of events in normalization mode ($B\to\bar{D}^{(*)}\ell^+\nu_{\ell}$) is extracted from the missing mass distribution 
in the region $-0.2 < M^2_{\rm miss} < 0.85$  $\rm GeV$\footnote{Natural units system with $\hslash$ = $c$ = $1$ is used.}, where 
$M^2_{\rm miss} = (p_{\rm beam} - p_{\rm B_{\rm tag}} - p_{D^{*}} - p_{l} )^2$ is the squared effective mass of neutrinos.

%
%

Backgrounds can be categorized into four components: $B \to \anti{D}^{*}\ell^+\nu_{\ell}$, $B \to \anti{D}^{**}\ell^+\nu_{\ell}$ together with hadronic $B$ decays, fake $D^*$ and continuum ($e^{+}e^{-} \to q\bar{q}$). 
The semileptonic component contaminates the signal sample due to the misassignment of the lepton as a pion, and it is fixed from the fit to the normalization sample.
In this analysis, the main background comes from hadronic $B$ decays with a few missing final-state particles, and its yield is determined as a free parameter in the final fit. 
The yield of the fake $D^*$ component is fixed from a comparison of the data and MC in the sideband regions of the mass diffrence between $D^*$ and $D$ mesons.
The fraction of the continuum process is negligible and is fixed using MC expectation.


%
Signal extraction is done by a 2D extended binned maximum likelihood fit to $E_{\rm ECL} $(summed energy of clusters not used in the reconstruction of $B_{\rm sig}$ and $B_{\rm tag}$ candidates) and $M^{2}_{\rm miss}$ distributions.                                                        
The fit is performed in two steps; the first fit is to the normalization sample, and then a simultaneous fit
for the eight signal samples: $(B^{+},B^{0}) \otimes (\pi^{+} \anti{\nu}_{\tau}, \rho^{+} \anti{\nu}_{\tau})\ \otimes$ (forward, backward).
The following result is obtained\cite{Hirose, Hirose2}:
\begin{align*}
P_{\tau}^{D^*} = -0{.}38 \pm 0{.}51(stat.) ^{+0{.}21}_{-0{.}16}(syst.) \\
R(D^*) = 0{.}270 \pm 0{.}035(stat.) ^{+0{.}028}_{-0{.}025}(syst.)
\end{align*}
Dominant systematics comes from hadronic $B$ decays composition ($^{+7{.}6\%}_{-6{.}8\%}$, $^{+0{.}13}_{-0{.}10}$)
and limited MC statistics for probability density functions
(PDFs) of shapes ($^{+4{.}0\%}_{-2{.}8\%}$, $^{+0{.}15}_{-0{.}11}$).
Combined $R(D^*)$ and $P_{\tau}^{D^*}$ result is consistent with the SM within $0.6\sigma$.
These are still crude constraints due to limited statistics but
better precision can be expected at Belle II experiment.

\subsection{$D^{*}$ polarization measurement in $B^0 \to D^{*-} \tau^{+} \nu_{\tau}$}

In the context of present experimental situation, where all $R(D^{(*)})$ measurements are systematically above the SM expectations, 
an interesting observable to study is the $D^*$ polarization in $B \to \anti{D}^{*} \tau^{+} \nu_{\tau}$ decays.
Moreover, by studying correlations amongst various observables, one can distinguish between the contributions from different NP operators~\cite{Bhattacharya:2018kig, Huang, Alok, Bhattacharya:2015ida, Ivanov}.

%
Fraction of the longitudinal polarization of $D^{*}$ is defined as: 
\begin{equation}
F_L^{D^*}=\frac{\Gamma (D^*_L)}{\Gamma(D^*_L)+\Gamma (D^*_T)}
\label{FL}
\end{equation}
where $\Gamma({D^{*}_{L(T)}})$ denote the decay rate with the longitudinally(transversely) polarised $D^*$, and $F_L^{D^*} + F_T^{D^*} = 1$. In SM,  $F_{L}^{D^*}$ is expected to be around 0.45. The most recent predictions are $F_{L,\rm SM}^{D^*} =0.457\pm 0.010$~\cite{Bhattacharya:2018kig}, $F_{L,\rm SM}^{D^*} =0.441\pm 0.006$~\cite{Huang}, and $F_{L,\rm SM}^{D^*} = 0{.}46\pm0{.}04$~\cite{Alok}.
$F_{L}^{D^*}$ can be significantly modified in the presence of NP contributions, in particular, the scalar (tensor) operators may enhance (decrease) $F_{L}^{D^*}$~\cite{Alok}.

In this report the first measurement of $D^*$ polarization in the $B^0 \to D^{*-} \tau^{+} \nu_{\tau}$ mode is presented.  
$F_{L}^{D^*}$ is extracted from angular distribution in $D^{*-} \to \anti{D}^0\pi^{-}$ decays. The polar angle distribution in the helicity frame is given by
\begin{equation}
\frac{1}{\Gamma} \frac{d\Gamma}{d \cos \theta_{\rm hel}^{D^*}} = \frac{3}{4} [2 F_L^{D^*} \cos^2\theta_{\rm hel}^{D^*}
  + F_T^{D^*} \sin^2\theta_{\rm hel}^{D^*}],
\label{cosThetaHelDstar}
\end{equation}\\
Compared to the $P_{\tau}$, $D^{*}$ polarization is easier to measure since all $\tau$ decays can be used, and it is not affected by cross-feeds between different $\tau$ decay channels. However, strong acceptance effects make measurements in the full $\cosDstar$ range difficult.
In particular, the region of $\cosDstar > 0$ is depleted due to the fact that at $\cosDstar$ close to +1, the pion goes backwards in the $D^*$ rest frame, and thus has lower momentum in the laboratory frame. The effect increases with higher value of $q^2$. Nevertheless, physical $\cosDstar$ distribution is symmetric, so 
range $-1 \leq \cosDstar \leq 0.0$ is sufficient to measure the $D^{*}$ polarization. 

In case of this analysis inclusive $B_{\rm tag}$ reconstrucion method was employed. Signal side was reconstructed using decay chains that combine a high reconstruction efficiency with a low background level:
$B^0 \to D^{*-} \tau^{+} \nu_{\tau}$,\\ 
$D^{*-}\to \anti{D}^{0} \pi^{-}$,\ $\anti{D}^{0} \to K^{+}\pi^{-},\ K^{+}\pi^{-}\pi^0,\ K^{+}\pi^{-}\pi^{+}\pi^{-}$,\ $\tau^{+} \to e^{+}\nu_{e}\anti{\nu}_{\tau},\ \mu^{+} \nu_{\mu}\anti{\nu}_{\tau},\ \pi^{+}\anti{\nu}_{\tau}$. 
The consistency of a $B_{\rm tag}$ candidate with a $B$ meson decay is checked using the beam-energy constrained mass and the energy difference variables: 
\begin{equation}
\begin{split}
M_{\rm tag} = \sqrt{E^{2}_{\rm beam} - \vec{p}^{2}_{\rm tag}}, \quad \vec{p}_{\rm tag} = \sum_i \vec{p}_{i} \\ 
\Delta E_{\rm tag} = E_{\rm beam} - E_{\rm tag}, \quad E_{\rm tag} = \sum_i E_{i} 
\end{split}
\end{equation}
where $\vec{p}_{i}$ and $E_{i}$ denote the 3-momentum vector and energy of the particle $i$ assigned to $B_{\rm tag}$ in the $\Upsilon(\rm 4S)$ rest frame, respectively.
The main backgrounds come from two other semileptonic $B$ decays ($B \to \anti{D}^{*}\ell^+\nu_{\ell}$, $B \to \anti{D}^{**}\ell^+\nu_{\ell}$).
For the background suppresion, the observables sensitive to multiple neutrino final states, like visible energy and
variable that approximates missing mass $X_{\rm mis} = ( E_{\rm beam}-E_{D^{*}}-E_{d} - |\vec{p}_{D^{*}} + \vec{p}_{d}| ) / ( \sqrt{E^{2}_{beam} - m^{2}_{B}} )$, but does not depend on tagging side, were used. $E_{d}$, $\vec{p}_{d}$ denote energy and momentum of the $\tau$ daughter respectively, and $m_{B}$ is the nominal mass of $B$ meson.

\begin{figure}[h]
  \centering
\includegraphics[width=0.475\textwidth]{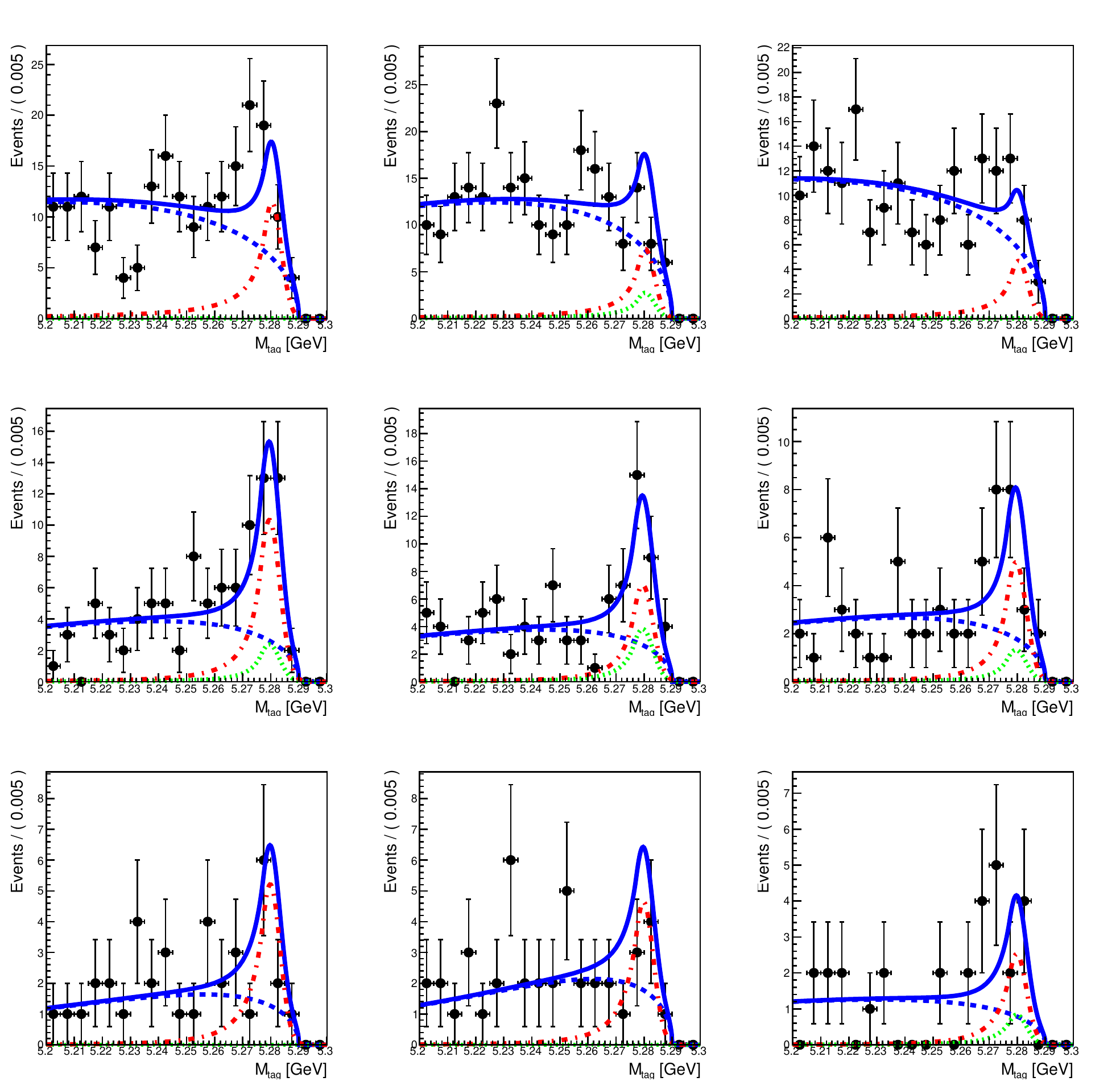} 
\caption{
Fit projection to $M_{\rm tag}$ distributions in the range $-1 \leq \cosDstar <  0.67$ for $\tau \to \pi \nu_{\tau}$ (top), for $\tau \to e \anti{\nu}_{e} \nu_{\tau}$ (middle), for $\tau \to \mu \anti{\nu}_{\mu} \nu_{\tau}$ (bottom), and $D^0 \to K\pi$ (left), $D^0 \to K\pi\pi^0$ (center), $D^0 \to K\pi\pi\pi$ (right). Data are points with error bars, and the fit results are represented by the solid blue lines. Contributions from signal, combinatorial and peaking backgrounds are shown by the red (dot-dashed), blue (dashed) and green (dotted) lines, respectively.   
}
\label{fit_projection}
\end{figure}

\begin{figure}[h]
  \centering
\includegraphics[width=.325\textwidth]{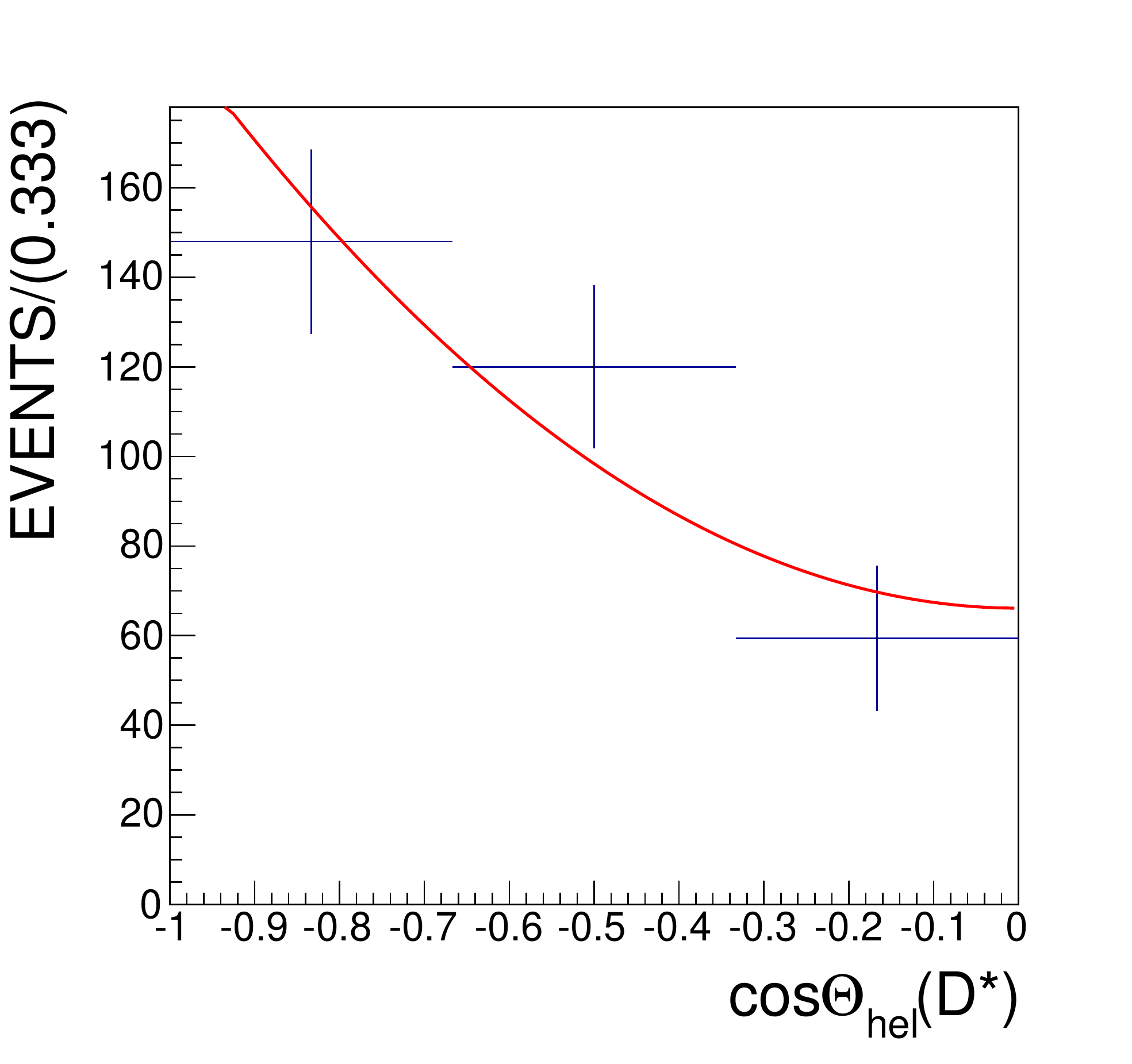}
\caption{
The measured $\cosDstar$ distribution for $B^0 \to D^{*-} \tau^{+} \nu_{\tau}$ mode. The red line represents the fit result with $F_L^{D^*} = 0{.}60\pm0{.}08(stat.)$.
}
\label{FL_fit}
\end{figure}

The signal yields are extracted using simultaneous, extended unbinned maximum likelihood fits to the $M_{\rm tag}$ distribution, where the sample was splitted up into three decays modes for $D^0$ and $\tau$, giving 9 sub-samples total. 
In this approach signal is extracted using known PDF's parametrizations, namely CrystalBall function for singal and peaking background, and Argus function for combinatorial background.
These fits are performed in 3 bins of $\cosDstar$ in the range of $-1 < \cosDstar < 0 $, where I bin: $-1.0 <$ $\cosDstar$ $< -0.67$, II bin: $-0.67 <$ $\cosDstar$ $< -0.33$, III bin: $-0.33 <$ $\cosDstar$ $< 0.0$. 
An example of the fit projection to $M_{\rm tag}$ distribution in the range $-1.0 < \cosDstar < -0.67$ is illustrated in Fig.\ref{fit_projection}. 
%
%
%
Following yields in the three bins of $\cosDstar$ were obtained: $N_{\rm I} = 151 \pm 21$, $N_{\rm II} = 125 \pm 19$, $N_{\rm III} = 55 \pm 15$. The corresponding statistical significances ($\Sigma$) in each bins are $\Sigma_{\rm I} = 8.8$, $\Sigma_{\rm II} = 7.8$, $\Sigma_{\rm III} = 4.1$. The statistical significance is defined as $\Sigma = \sqrt{-2 ln(\mathcal{L}_{0}/\mathcal{L}_{\rm max})}$, where $\mathcal{L}_{\rm max}$ and $\mathcal{L}_{\rm 0}$ denote the maximum likelihood value and the likelihood value for the zero signal hypothesis. 
Additionally, taking into account small acceptance variations and bin migration, the signal yields in the bins of $\cosDstar$ are reweighted with the following scale factors: $s_{\rm I} = 0.98 \pm 0.01$, $s_{\rm II} = 0.96 \pm 0.01$, $s_{\rm III} = 1.08 \pm 0.01$.    
Fig.\ref{FL_fit} shows the measured $\cosDstar$ distribution, 
which is fitted using the formula (\ref{cosThetaHelDstar}) to extract $F_L^{D^*}$. The fit result gives 
$F_L^{D^*} = 0{.}60\pm0{.}08(stat.)\pm0{.}04(syst.)$, and agrees within 1.6, 1.8 and 1.4 standard deviation with the SM predictions $F_{L,\rm SM}^{D^*} =0.457\pm 0.010$~\cite{Bhattacharya:2018kig}, $F_{L,\rm SM}^{D^*} = 0.441\pm 0.006$~\cite{Huang}, and $F_{L,\rm SM}^{D^*} = 0{.}46\pm0{.}04$~\cite{Alok}, respectively.
The dominant systematic uncertainties arise from the limited MC statistics for determination of background PDF shapes ($\pm0{.}03$) and imperfect modelling of real processes ($B \to D^{**} \ell \nu$) ($\pm0{.}01$). 
The full potential of data sample can be further exploited by adding charged $B$ mode.
%


\section{Prospects for Belle/Belle II}

Belle is still active in carrying out new measurements using newly developed analysis tools. In recent analyses it was decided 
that Belle data are converted to Belle II data format~\cite{B2BII}, what enables to use Belle II software framework (BASF2). In particular, to reconstruct $B_{\rm tag}$ using hadronic or semileptonic $B$ decays we can exploit Full Event Interpretation (FEI)~\cite{FEI} - a new exclusive tagging algorithm in BASF2 for multivariate analysis with Boosted-Decision Tree (BDT) classifier. 
The new tool allows for increasing the signal reconstruction efficiency by the factor of three (without increasing the background) compared to the previous algorithm (see e.g. \cite{Gelb})
FEI is used in the currently ongoing analysis on simultaneous measurement of $R(D)$ and $R(D^*)$ tagged with semileptonic $B$ decays.

\begin{figure}
  
\includegraphics[width=.415\textwidth]{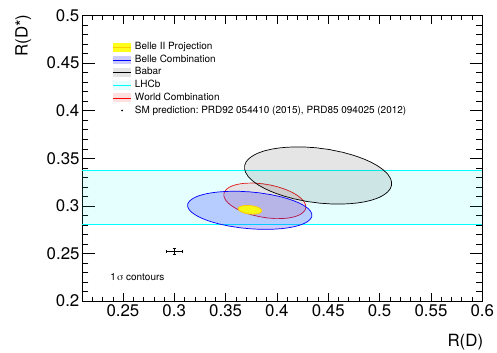}
\includegraphics[width=.415\textwidth]{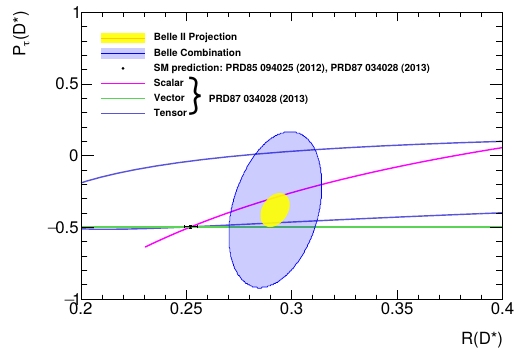}
\caption{
Expected constraints on $R(D)$ vs. $R(D^*)$ (top pannel) and $P^{D^{*}}_{\tau}$ vs. $R(D^*)$ (bottom pannel) compared to existing experimental constraints.
Current values with uncertainties of $R(D^{(*)})$ and $P^{D^{*}}_{\tau}$, and projection for 50 ab$^{-1}$ are based on Ref.\cite{BelleIIBook}.
}
\label{RD_RDstar_plane}
\end{figure}

Data taking with the Belle II detector is expected to kick-off in early 2019 and 
collect approximately 50 times Belle data sample by 2025. 
Present and expected precision for $R(D^{(*)})$, $P_{\tau}^{D^*}$ and $F_L^{D^*}$ with 5ab$^{-1}$ data sample and the full ~50ab$^{-1}$ data are summarized in Table~\ref{tab:expected_precision} (based on Ref.\cite{BelleIIBook}). Comparison between current and expected precision is illustrated in Fig.\ref{RD_RDstar_plane}. 
The ultimate relative uncertainty (with ~50ab$^{-1}$) of 2\% for $R(D)$ and 1\% for $R(D^{*})$ would be dominated by systematics.
The major contribution to $R(D^{(*)})$ systematics is the uncertainty on branching fraction of $D^{**}$ components.
The $B \to \bar{D}^{**} X$ modes, especially $B \to \bar{D}^{**} \ell \nu$, will be studied in detail, so a simultanous determination of $R(D)$, $R(D^{*})$ and $R(D^{**})$ should be possible in the long term.   
Next source of systematics are multibody hadronic $B$ decays, especially with neutral particles in final states.
However, our understanding of these modes should be improved by future measurements at Belle II and hence the systematics uncertainty can be reduced. 

Using inclusive $B_{\rm tag}$ reconstruction method we can expect at least 15000 events for the $F_{L}^{D^*}$ determination. The estimation  assumes comparable tagging performance and reconstruction efficiency as in Belle. 
Higher statistics and better reconstrucion efficiencies at Belle II should allow for precise test of NP scenarios with $q^2$, and other differential distributions of kinematic observables. 

\begin{table}
\centering
\begin{tabular}{ c c c }
    \hline
        & 5 ab$^{-1}$ & 50 ab$^{-1}$ \\
		\hline 
		$R(D)$ & ($\pm 6\pm4$)\% & ($\pm 2\pm3$)\%~\cite{BelleIIBook}\\
		\hline
		$R(D^*)$ & ($\pm 3\pm3$)\% & ($\pm 1\pm2$)\%~\cite{BelleIIBook}\\
		\hline
        $P_{\tau}^{D^*}$ & $\pm 0.18\pm0.08$ & $\pm 0.06\pm0.04$~\cite{BelleIIBook}\\
        \hline
        $F_L^{D^*}$ & $\pm0.04\pm0.04$ & $\pm0.01\pm0.04$\\
		\hline
	\end{tabular}
\caption{Expected precision for $R(D^{(*)})$ and $P_{\tau}^{D^*}$ at Belle II, given as the relative uncertainty for $R(D^{(*)})$ and absolute ones for $P_{\tau}^{D^*}$. The values given are the statistical and systematic errors respectively.}
\label{tab:expected_precision}
\end{table}

\section{Summary}

In this review the current experimental status for semitauonic $B$ decays and first meaurement of $D^*$ polarization in $B^0 \to D^{*-} \tau^{+} \nu_{\tau}$  are reported. The fraction of $D^{*-}$ longitudinal polarization, measured assuming the SM dynamics, is found to be $F_L^{D^*} = 0{.}60\pm0{.}08(stat.)\pm0{.}04(syst.)$, and agrees within 1.6, 1.8 and 1.4 standard deviation with the SM predictions $F_{L,\rm SM}^{D^*} =0.457\pm 0.010$~\cite{Bhattacharya:2018kig}, $F_{L,\rm SM}^{D^*} = 0.441\pm 0.006$~\cite{Huang}, and $F_{L,\rm SM}^{D^*} = 0{.}46\pm0{.}04$~\cite{Alok}, respectively.

At present, sensitivity of these measurements is limited by statistics, and exploration of characteristics of semitauonic $B$ decays will be important topic at the Belle II experiment.

\subsection*{Acknowledgements}
I would like to thank the hosts of CKM 2018 workshop for the opportunity to give this talk at such well organized and interesting conference. This research is supported by the Polish Ministry of Science and Higher Education through grant DIR/WK/2017/03.

\end{document}